\documentclass[11p,a4]{article}   

\usepackage{amssymb}
\usepackage{amsmath}
\usepackage{graphicx}
\usepackage[numbers,compress]{natbib}
\begin{document}

\title{Flat limit of the de Sitter QFT  in the rest frame vacuum}

\author{Ion I. Cot\u aescu 
\thanks{e-mail: i.cotaescu@e-uvt.ro}\\
{\em West University of Timi\c soara, V. P\^ arvan Ave. 4,}\\ {\em RO-300223, Timi\c soara, Romania}}

\maketitle

\begin{abstract}
The problem of the flat limits of the scalar and spinor fields on the de Sitter expanding universe is considered in the rest frame vacuum where the frequencies are separated  in rest frames as in special relativity. The  phases of the fundamental solutions of these fields are regularized in order to obtain  Minkowskian  flat limits. New expressions of  scalar mode functions or Dirac spinors reaching natural flat limits are given  in the rest frame vacuum 
and suitable approximate formulas are proposed. In this manner the flat limit  is correctly defined for the entire quantum theory since the one-particle operators tend to those of special relativity.  
 
Pacs: 04.62.+v
\end{abstract}

\section{Introduction}

The quantum field theory (QFT) on the expanding portion of the de Sitter space-time can be studied extensively since the equations of the principal free fields can be solved analytically in various local charts. In the co-moving charts with Cartesian space coordinates there are plane waves solutions which are eigenfunctions of the momentum operator, as in special relativity, but with more complicated time modulation functions. Moreover, in this geometry one cannot use the Hamiltonian operator for separating the frequencies since this does not commute with the momentum operator. Thus the main  task here is just the criterion of separating the frequencies, defining the particle and antiparticle modes and, implicitly, the vacuum \cite{BD}.   

The principal method used so far focused mainly on the asymptotic states which are somewhat similar  to  the usual Minkowskian particle or antiparticle states, as in the case of the adiabatic vacuum of the Bunch-Davies type \cite{BuD} largely used in applications. The problem arising in this vacuum  is that one cannot reach the rest limit, for vanishing momentum, since this limit is undefined for the corresponding scalar mode functions or  Dirac spinors. This problem was considered sporadically by few authors  which found similar  results \cite{nach,born,CPasc,CD}. 

Under such circumstances,  we proposed recently a method of separating the frequencies in  rest frames defining thus the rest frame vacuum (r.f.v.) of the massive Klein-Gordon \cite{CrfvKG}, Dirac \cite{CrfvD} and Proca \cite{CrfvP} fields.  What is new here is that the bosons can have either a tardyionic behavior or even a tachyonic one if the mass is less than a given limit which depends on the type of coupling (minimal or conformal). Fortunately, the tachyonic modes are eliminated in a natural manner since all their mode functions have null norms \cite{CrfvKG}. In contrast, the Dirac field of any non-vanishing mass survives in this vacuum \cite{CrfvD}. In other respects, we must specify the r.f.v. can be defined only for massive particles since the massless ones do not have rest frames. This is not an impediment since the massless fields of physical interest, namely the Maxwell and neutrino ones, have conformally covariant  field equations whose solutions in the co-moving de Sitter chart with conformal time can be taken over from special relativity \cite{Max,CD1}.     

Technically speaking, for defining the  r.f.v. we introduced suitable phases depending on momentum for assuring the correct limits of the mode functions in rest frames \cite{CrfvKG,CrfvD,CrfvP}. Unfortunately, these phases are not enough for defining the other important limit, namely the flat one, when the de Sitter Hubble constant tends to zero. In general, this limit is undefined because of  some singularities arising in  the phases of the mode functions defined in the adiabatic vacuum. For removing them a regularization procedure was applied by adding the convenient phase factors which, in general, depend on momentum  \cite{nach,born,CPasc,CD}. 

Since this problem was not yet considered for the recently defined r.f.v., the aim of  this paper is to study the flat limits of the Klein-Gordon and Dirac fundamental solution in this vacuum, deriving the regularized phases which guarantee that these  limits are well-defined. In this manner,  the rest and flat limits determine completely the form of the mode functions or spinors of the de Sitter QFT .   This is important since there are quantities whose expressions are strongly dependent on the  momentum-dependent phase factors as, for example, the one-particle Hamiltonian operator \cite{nach,CD1,Cpop}.  We prove that the phase factors derived here determine the correct Minkowskian flat limit of this operator.   

We must specify that in what concerns the regularized phases we recover previous results \cite{nach,born,CPasc,CD}  but the complete expressions of the scalar mode functions or Dirac spinors are presented here for the first time. Moreover, the approximations we obtain here are also new results that can be used in concrete calculations.

We start  presenting in the next section the plane wave fundamental solutions of the Klein-Gordon and Dirac fields in the adiabatic and rest frame vacua pointing out how the last one solves the rest limits. The third section is devoted to the flat limits which are derived by using a new uniform asymptotic expansion we propose here based on numerical arguments. This enables us to derive the regularized phases which assure the flat limits of the fundamental solutions. Thus we obtain for the first time  the complete expressions of the scalar mode  functions and Dirac spinors in the r.f.v. as well as useful approximations of them.  Moreover, we show that in this approach the flat limit of the one-particle Hamiltonian operator is just the corresponding operator of the Minkowskian QFT. Finally we present our concluding remarks.   

\section{Free fields in adiabatic and rest frame vacua}

The  $(1+3)$-dimensional de Sitter expanding universe, $M$,  is the expanding portion of the de Sitter space-time where we may choose the co-moving local charts $\{t,{\bf x}\}$ whose coordinates $x^{\kappa}$ (labelled  by the natural indices $\kappa,\nu,...=0,1,2,3 $) are the proper (or cosmic) time, $t$, and the Cartesian space coordinates $x^i$  ($i,j,k...=1,2,3$)  for which we may use the vector notation ${\bf x}=(x^1,x^2,x^3)$.  The geometry is given by the scale factor $a(t)=\exp(\omega t)$ depending on the Hubble de Sitter parameter denoted here by $\omega$. Another useful chart is that of the conformal time,
\begin{equation}\label{tct}
t_c=\int \frac{dt}{a(t)}=-\frac{1}{\omega}e^{-\omega t}~\to~  a(t_c)=a[t(t_c)]=-\frac{1}{\omega t_c}\,,
\end{equation}
and the same Cartesian space coordinates, denoted by $\{t_c,\vec{x}\}$. The line elements of these charts are,  
\begin{eqnarray}
ds^2=g_{\kappa\nu}(x)dx^{\kappa}dx^{\nu}&=&dt^2-e^{2\omega t} d{\vec x}\cdot d{\vec x}\nonumber\\
&=&\frac{1}{(\omega t)^2}(dt_c^2-d{\vec x}\cdot d{\vec x})\,,
\end{eqnarray}
Note that on the expanding portion we have $t_c\in (-\infty,0]$ and $t\in(-\infty,\infty)$.

For writing down the Dirac equation  we chose the diagonal tetrad gauge in which the vector fields $e_{\hat\alpha}=e_{\hat\alpha}^{\kappa}\partial_{\kappa}$ defining the local orthogonal frames,  and the 1-forms $\omega^{\hat\alpha}=\hat e_{\kappa}^{\hat\alpha}dx^{\kappa}$ of the dual coframes (labeled by the local indices, $\hat\kappa,\hat\nu,...=0,1,2,3$) are defined as 
\begin{eqnarray}
&e_0=\partial_t=\frac{1}{a(t_c)}\,\partial_{t_c}\,,\qquad & \omega^0=dt=a(t_c)dt_c\,,\label{tetrad}\label{T1} \\
&~~~e_i=\frac{1}{a(t)}\,\partial_i=\frac{1}{a(t_c)}\,\partial_i\,, \qquad & \omega^i=a(t)dx^i=a(t_c)dx^i\,,\label{T2}
\end{eqnarray}
in order to preserve the global $SO(3)$ symmetry  allowing us to use systematically the  $SO(3)$ vectors. We remind the reader that the metric tensor of $M$ can be expressed now as $g_{\kappa\nu}=\eta_{\hat\alpha\hat\beta}\hat e^{\hat\alpha}_{\kappa}\hat e^{\hat\beta}_{\nu}$ where $\eta={\rm diag}(1,-1,-1,-1)$ is the Minkowski metric.

\subsection{Klein-Gordon field}

In the chart $\{t,\vec{x}\}$ the scalar field $\Phi : M\to {\Bbb C}$ of mass $m$, minimally coupled to the de Sitter gravity, satisfies the Klein-Gordon equation,
\begin{equation}\label{KG1}
\left( \partial_t^2-e^{-2\omega t}\Delta +3\omega
\partial_t+m^2\right)\Phi(x)=0\,,
\end{equation}
whose general solutions can be expanded as
\begin{equation}\label{field1}
\Phi(x)=\Phi^{(+)}(x)+\Phi^{(-)}(x)=\int d^3p \left[f_{\vec{p}}(x)a(\vec{p})+f_{\vec{p}}^*(x)b^{\dagger}(\vec{p})\right] \,,
\end{equation}
in terms of  field operators, $a(\vec{p})$ and $b(\vec{p})$,  and fundamental solutions, $f_{\vec{p}}$ and  $f_{\vec{p}}^*(x)$, of positive  and respectively negative frequencies. These solutions must satisfy the orthonormalization relations
\begin{eqnarray}
\langle  f_{\vec{p}},f_{\vec{p}'}\rangle_{KG}=-\langle  f_{\vec{p}}^*,f_{\vec{
p}'}^*\rangle_{KG}&=&\delta^3(\vec{p}-\vec{p}')\,,\label{ff}\\
\langle f_{\vec{p}},f_{\vec{p}'}^*\rangle_{KG}&=&0\,,
\end{eqnarray}
and a completeness condition with respect to the relativistic scalar product \cite{BD}
\begin{equation}\label{SP}
\langle f,f'\rangle_{KG}=i\int d^3x\, a(t)^3\, f^*(x)
\stackrel{\leftrightarrow}{\partial_{t}} f'(x)\,.
\end{equation}
The fundamental mode functions can be expressed as  
\begin{equation}\label{fp}
f_{\vec{p}}(t,\vec{x})=\frac{e^{i \vec{x}\cdot \vec{p}}}{[2\pi a(t)]^{\frac{3}{2}}}{\cal F}_p(t)\,,
\end{equation}
in terms of the time modulation functions  ${\cal F}_p: D_t\to {\Bbb C}$  which depend on $p=|\vec{p}|$ satisfying the equation
\begin{equation}\label{KGred}
\left[\frac{d^2}{dt^2}+\frac{p^2}{a(t)^2}+m^2-\frac{9}{4}\,\omega^2\right] {\cal F}_p(t)=0\,.
\end{equation}
and the normalization condition 
\begin{equation}\label{normF}
\left({\cal F}_p, {\cal F}_p\right)\equiv  i\,{\cal F}_p^*(t)\stackrel{\leftrightarrow}{\partial}_{t}{\cal F}_p(t)=1\,.
\end{equation}
 which guarantees the condition  (\ref{ff}).

The most general solution of  Eq. (\ref{KGred}) can be derived easily in the chart $\{t_c,\vec{x}\}$ obtaining \cite{Cpop,CrfvKG}
\begin{equation}\label{fdS}
{\cal F}_p(t_c)=\ c_1\phi_p(t_c) + c_2\phi_p^*(t_c)\,,\quad \phi_p(t_c)=\frac{1}{\sqrt{\pi\omega}}\,K_{\nu}(ipt_c)\,,
\end{equation}
where $K$ is the modified Bessel function of the index
\begin{equation}\label{ndS}
\nu=\left\{\begin{array}{lll}
\sqrt{\frac{9}{4}-\mu^2}&{\rm for} & \mu<\frac{3}{2}\\
i\kappa\,,\quad \kappa= \sqrt{\mu^2-\frac{9}{4}}&{\rm  for} & \mu>\frac{3}{2}
\end{array} \right. \,, \quad \mu=\frac{m}{\omega}\,.
\end{equation}
The particular solution $\phi_p(t_c)$ is normalized, satisfying $(\phi_p,\phi_p)=1$, such that  the condition (\ref{normF}) is fulfilled only if we take
\begin{equation}\label{norC}
\left|c_1\right|^2-\left|c_2\right|^2=1\,.
\end{equation}
Thus we remain with an undetermined integration constant that may be fixed by giving a criterion of frequencies separation setting thus the vacuum. 

The most popular vacuum is the adiabatic Bunch-Davies one \cite{BuD}, with $c_1=1$ and $c_2=0$,  that holds for any mass, regardless  the real or imaginary value of the index (\ref{ndS}). Despite of this advantage here we face with the problem of the rest limit which cannot be defined as long as the functions $K_{i\kappa} (ipt_c)$ have an ambiguous  behavior,
\begin{equation}
 K_{i\kappa} (ipt_c)\propto \frac{1}{\Gamma(\frac{1}{2}-i\kappa)}\left(\frac{ipt_c}{2}\right)^{-i\kappa}-\frac{1}{\Gamma(\frac{1}{2}+i\kappa)}\left(\frac{ipt_c}{2}\right)^{i\kappa}\,,
\end{equation}
for $p\to 0$, as it results from Eq. (\ref{I0}). A  possible solution is to redefine these functions replacing $i\kappa\to i\kappa \pm \epsilon$ for eliminating one of the above terms and introducing a convenient phase factor for the remaining one \cite{nach,born,CPasc}. However,  this procedure is palliative since this affects the physical meaning of the mass which gets an imaginary part.

For avoiding these difficulties we defined recently the r.f.v., separating the frequencies in the rest frames just as in special relativity \cite{CrfvKG}. Thus we found that the rest energy,
\begin{equation}\label{Mko}
M=\kappa\omega=\sqrt{m^2-\frac{9}{4}\omega^2}\,, \quad m>\frac{3}{2}\omega\,,
\end{equation}
which plays the role of a dynamical mass, does make sense only for  $\mu>\frac{3}{2}$  since for $\mu<\frac{3}{2}$ the mode functions do not have a physical meaning being of tachyonic type but with null norms.  We have shown that in the tardyonic domain this vacuum is stable corresponding to the integration constants,
\begin{equation}
c_1=-i\left(\frac{p}{2\omega}\right)^{-i\kappa}\frac{e^{\pi\kappa}}{\sqrt{e^{2\pi\kappa}-1}}\,, \quad
c_2= i\left(\frac{p}{2\omega}\right)^{-i\kappa}\frac{1}{\sqrt{e^{2\pi\kappa}-1}}\,,
\end{equation} 
determining the  time modulation functions of positive energy as,  \cite{CrfvKG}
\begin{equation}\label{Ftc}
{\cal F}_p(t_c)=\sqrt{\frac{\pi}{\omega}}\left(\frac{p}{2\omega}\right)^{-i\kappa}\frac{I_{i\kappa}(ipt_c)}{\sqrt{e^{2\pi\kappa}-1}}\,.
\end{equation}
We must specify that the above phase factor is introduced for assuring the correct rest limit 
\begin{equation}
\lim_{p\to 0} {\cal F}_p(t_c) =\frac{1}{\sqrt{2M}}\,e^{-i M t}\,,
\end{equation}
calculated according to Eqs. (\ref{tct}) and  (\ref{I0}).   

\subsection{Dirac field}

In the tetrad-gauge defined by Eqs. (\ref{T1}) and (\ref{T2}) the massive Dirac field $\psi$ of mass $m$  satisfies the field equations $(D_x-m) \psi (x)=0$ given by the Dirac operator 
\begin{equation}\label{ED}
D_x=i\gamma^0\partial_{t}+i e^{-\omega t}\gamma^i\partial_i
+\frac{3i \omega}{2} \gamma^{0}\,.
\end{equation}
where  $\gamma^{\hat\alpha}$ are the Dirac matrices labeled by local indices.  It is known that the last term of this operator  can be removed at any time by substituting $\psi \to [a(t)]^{-\frac{3}{2}}\psi$. Similar results can be written in the conformal chart.

The general solution of the Dirac equation  may be written as a mode integral, 
\begin{eqnarray}
\psi(t,{\bf x}\,)& =& 
\psi^{(+)}(t,{\bf x}\,)+\psi^{(-)}(t,{\bf x}\,)\nonumber\\
& =& \int d^{3}p
\sum_{\sigma}[U_{\vec{p},\sigma}(x){\frak a}(\vec{p},\sigma)
+V_{\vec{p},\sigma}(x){\frak b}^{\dagger}(\vec{p},\sigma)]\,,\label{p3}
\end{eqnarray}
in terms of the fundamental spinors $U_{\vec{p},\sigma}$  and  $V_{\vec{p},\sigma}$ of positive and respectively negative frequencies which are plane waves solutions of the Dirac equation depending on the conserved momentum $\vec{p}$ and an arbitrary polarization $\sigma$. These spinors  form an orthonormal  basis being related  through the charge conjugation, 
\begin{equation}\label{chc}
V_{\vec{p},\sigma}(t,{\bf x})=U^c_{\vec{p},\sigma}(t,{\bf x}) =C\left[{U}_{\vec{p},\sigma}(t,{\bf x})\right]^* \,, \quad C=i\gamma^2\,,
\end{equation}
(see the  Appendix A), and satisfying the orthogonality relations
\begin{eqnarray}
\langle U_{\vec{p},\sigma}, U_{{\vec{p}\,}',\sigma'}\rangle_D &=&
\langle V_{\vec{p},\sigma}, V_{{\vec{p}\,}',\sigma'}\rangle_D=
\delta_{\sigma\sigma^{\prime}}\delta^{3}(\vec{p}-\vec{p}\,^{\prime})\label{ortU}\\
\langle U_{\vec{p},\sigma}, V_{{\vec{p}\,}',\sigma'}\rangle _D&=&
\langle V_{\vec{p},\sigma}, U_{{\vec{p}\,}',\sigma'}\rangle_D =0\,, \label{ortV}
\end{eqnarray}
with respect to the relativistic scalar product \cite{CD1}
\begin{equation}
\langle \psi, \psi'\rangle_D=\int d^{3}x
\sqrt{|g|}\,e^0_0\,\bar{\psi}(x)\gamma^{0}\psi(x) 
=\int d^{3}x\,
a(t)^{3}\bar{\psi}(x)\gamma^{0}\psi(x)\,, 
\end{equation}
where $g={\rm det}(g_{\kappa\nu})$ and $\bar{\psi}=\psi^+\gamma^0$ is the Dirac adjoint of $\psi$. Moreover,  this basis is supposed to be complete complying with a completeness condition  \cite{CD1}. This is the basis of the momentum representation  in which   the particle $({\frak a},{\frak a}^{\dagger})$ and antiparticle (${\frak b},{\frak b}^{\dagger})$ operators  satisfy the canonical anti-commutation relations \cite{CD1}. 

In the standard representation of the Dirac matrices (with diagonal $\gamma^0$) the general form of the fundamental spinors in momentum representation,   
\begin{eqnarray}
U_{\vec{p},\sigma}(t,\vec{x}\,)&=&\frac{e^{i\vec{p}\cdot\vec{x}}}{[2\pi a(t)]^{\frac{3}{2}}}\left(
\begin{array}{c}
u^+_p(t) \,
\xi_{\sigma}\\
u^-_p(t) \,
 \frac{{p}^i{\sigma}_i}{p}\,\xi_{\sigma}
\end{array}\right)
\label{Ups}\\
V_{\vec{p},\sigma}(t,\vec{x}\,)&=&\frac{e^{-i\vec{p}\cdot\vec{x}}}{[2\pi a(t)]^{\frac{3}{2}}} \left(
\begin{array}{c}
v^+_p(t)\,
\frac{{p}^i{\sigma}_i}{p}\,\eta_{\sigma}\\
v^-_p(t) \,\eta_{\sigma}
\end{array}\right)
\,,\label{Vps}
\end{eqnarray}
is determined by the modulation functions $u^{\pm}_p(t)$ and $v^{\pm}_p(t)$  that depend only on $t$ and  $p=|\vec{p}|$.   The Pauli spinors $\xi_{\sigma}$ and $\eta_{\sigma}= i\sigma_2 (\xi_{\sigma})^{*}$ have to be correctly normalized,  $\xi^+_{\sigma}\xi_{\sigma'}=\eta^+_{\sigma}\eta_{\sigma'}=\delta_{\sigma\sigma'}$,  satisfying a natural completeness equation.
 
The time modulation functions $u_p^{\pm}(t)$ and $v_p^{\pm}(t)$  must be related as \cite{CrfvD}
\begin{equation}\label{VU}
v_p^{\pm}=\left[u_p^{\mp}\right]^*\,,
\end{equation}
for assuring the charge conjugation symmetry  (\ref{chc}).  Moreover,   the normalization conditions
\begin{equation}
|u_p^+|^2+|u_p^-|^2=|v_p^+|^2+|v_p^-|^2 =1 \label{uuvv}\\
\end{equation}
guarantee that Eqs.  (\ref{ortU}) and (\ref{ortV}) are accomplished. 

The time modulation functions can be derived easily in the conformal chart  where we have to solve the system
\begin{eqnarray}
\left[i\partial_{t_c}\mp m\, a(t_c)\right]u_p^{\pm}(t_c)&=&{p}\,u_p^{\mp}(t_c)\,,\label{sy1c}\\
\left[i\partial_{t_c} \mp m\, a(t_c)\right]v_p^{\pm}(t_c)&=&-{p}\,v_p^{\mp}(t_c)\,,\label{sy2c}
\end{eqnarray}
obtaining the general form \cite{CrfvD}
\begin{eqnarray}
u^{+}_p(t_c)&=&\sqrt{-\frac{p t_c}{\pi}}\left[c_1 K_{\nu_{-}}\left(i p t_c\right)+c_2 K_{\nu_{-}}\left(-i p t_c\right)\right]\,,\label{coco1}\\
u^{-}_p(t_c)&=&\sqrt{-\frac{p t_c}{\pi}}\left[c_1 K_{\nu_{+}}\left(i p t_c\right)-c_2 K_{\nu_{+}}\left(-i p t_c\right)\right]\,,\label{coco2}
\end{eqnarray}
where the orders of the modified Bessel functions $K$ are $\nu_{\pm}=\frac{1}{2}\pm i \kappa$ after we re-define $ \kappa= \frac{m}{\omega}$. 

 The particular solutions of Eqs. (\ref{coco1}) and Eqs. (\ref{coco2}) are normalized and orthogonal to each other such that the normalization condition (\ref{uuvv}) is satisfied  if  
\begin{equation}
|c_1|^2+|c_2|^2=1\,.
\end{equation}
The functions $v_p^{\pm}$ result from Eq. (\ref{VU}). 

The adiabatic vacuum is defined simply by choosing $c_1=1$ and $c_2=0$ as in Refs. \cite{nach,CD1}. The major difficulty of this vacuum  is that in the momentum-spin representation we cannot reach the rest frame limit even though the functions $K$ have now defined limits for $p\to 0$. This is because of the term $\frac{\vec{p}\cdot\vec{\sigma}}{p}$ whose  limit is undefined \cite{CrfvD}. Moreover, if we force the limit vanishing  this term by hand then we affect the normalization \cite{CD,CD1}. 

The solution is to adopt the r.f.v. imposing the conditions \cite{CrfvD}
\begin{equation}\label{rfv}
\lim_{p\to 0} u^{-}_p(t)=\lim_{p\to 0} v^{+}_p(t)=0\,,
\end{equation}
which drop out the contribution of the mentioned terms in rest frames.  These conditions are accomplished only if we take
\begin{equation}\label{con}
c_1=\frac{e^{\pi\kappa}p^{-i\kappa}}{\sqrt{1+e^{2\pi \kappa}}}\,, \quad c_2=\frac{i\,p^{-i\kappa}}{\sqrt{1+e^{2\pi \kappa}}}\,,
\end{equation}
determining  the definitive form of the modulation functions of positive frequencies as \cite{CrfvD}
\begin{equation}\label{uuI}
u_p^{\pm}(t_c)=\pm \frac{\sqrt{-\pi t_c}\, p^{\nu_-}}{\sqrt{1+e^{2\pi\kappa}}}\, I_{\mp\nu_{\mp}}(ipt_c)\,.
\end{equation}
The modulation functions of the negative frequencies have to be calculated according to Eq.(\ref{VU}). Thus we obtain fundamental spinors whose rest limits
\begin{eqnarray}
\lim_{\vec{p}\to 0} U_{\vec{p},\sigma}(t,{\bf x})&=&\frac{e^{-i mt}}{[2\pi a(t)]^{\frac{3}{2}}}\left(
\begin{array}{c}
\xi_{\sigma}\\
0
\end{array}\right)\,,\label{Ur}\\
\lim_{\vec{p}\to 0}V_{\vec{p},\sigma}(t,{\bf x})&=&\frac{e^{i mt}}{[2\pi a(t)]^{\frac{3}{2}}}\left(
\begin{array}{c}
0\\
\eta_{\sigma}
\end{array}\right)\,,\label{Vr}
\end{eqnarray} 
indicate that the rest energy of the Dirac field is $m$ just as in special relativity.

\section{Flat limits}

The time modulation functions studied above depend on the variable 
\begin{equation}\label{var}
x=-\omega t_c=\frac{p}{\omega}e^{-\omega t} 
\end{equation}
which in the rest limit tends to $0$ but in the flat limit, when $\omega \to 0$ and $-\omega t_c \to 1$, this tends to infinity. Therefore, for analyzing the behavior of the time modulation functions in the flat limit  we need to use an uniform expansion of the Bessel function $J_{i\kappa+\lambda}(\kappa x)$  for large values of $\kappa>0$, any $x>0$ and $\lambda=0,\pm\frac{1}{2}$. Unfortunately, we have a rigorous proof only for $\lambda=0$ such that we are forced to generalize this case based on some analytical and numerical arguments.

\subsection{Approximating method}

{ \begin{figure}
\centering
  \includegraphics[scale=0.37]{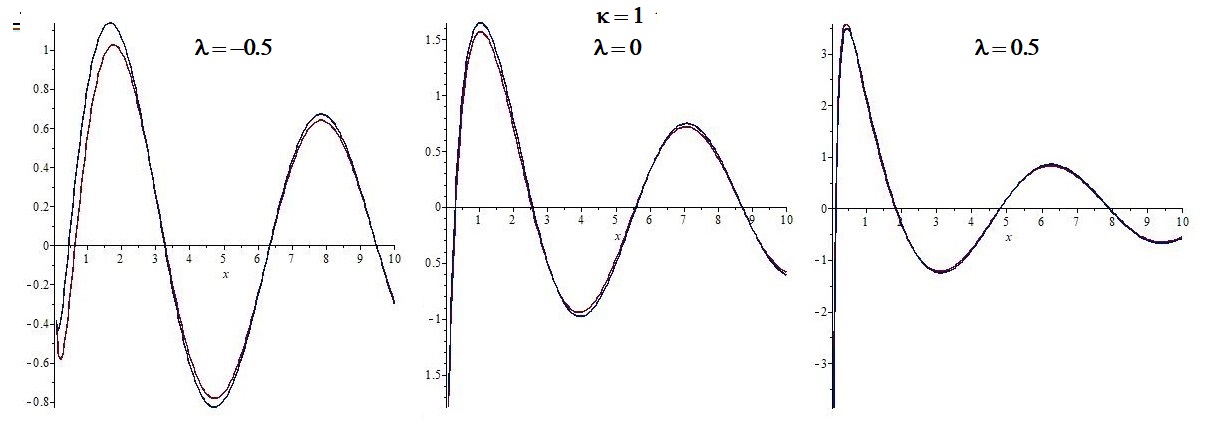}
  \caption{The functions $\Re\,J_{i\kappa+\lambda}(\kappa x)$ (red) and $\Re\,{\cal J}(\kappa,\lambda, x)$} (blue)  for $\kappa=1$ and $\lambda=-\frac{1}{2},0, \frac{1}{2}$.
 \end{figure}}

We propose a generalization of the standard uniform asymptotic expansion (\ref{unexA}) to $ J_{i\kappa+\lambda}(\kappa x)$ observing that this is analytic in $\kappa$ such that we can replace $i\kappa \to i\kappa+\lambda$   but without affecting  the variable $x$ or expressions containing it, as $\kappa x$ or $\kappa\sqrt{1+x^2}$. Therefore, we assume that the following approximation,
\begin{equation}\label{unex0}
J_{i\kappa+\lambda}(\kappa x) \simeq {\cal J}(\kappa,\lambda, x)=\frac{e^{\frac{\pi\kappa}{2}-\frac{i\pi\lambda}{2}}}{\sqrt{2\kappa\pi}}\,\frac{e^{i \kappa\sqrt{1+x^2}-\frac{i\pi}{4}}}{ (1+x^2)^{\frac{1}{4}}}
\left(\frac{1}{x}+\sqrt{1+\frac{1}{x^2}}\right)^{-i\kappa-\lambda}\,,
\end{equation}
in which we neglected the terms of the order ${\cal O}(\kappa^{-1})$, holds even for non-vanishing values of $\lambda\in {\Bbb R}$. However, the crucial point is to verify if this approximation is numerically satisfactory, comparing the functions $J$ and ${\cal J}$. 

{ \begin{figure}
\centering
  \includegraphics[scale=0.37]{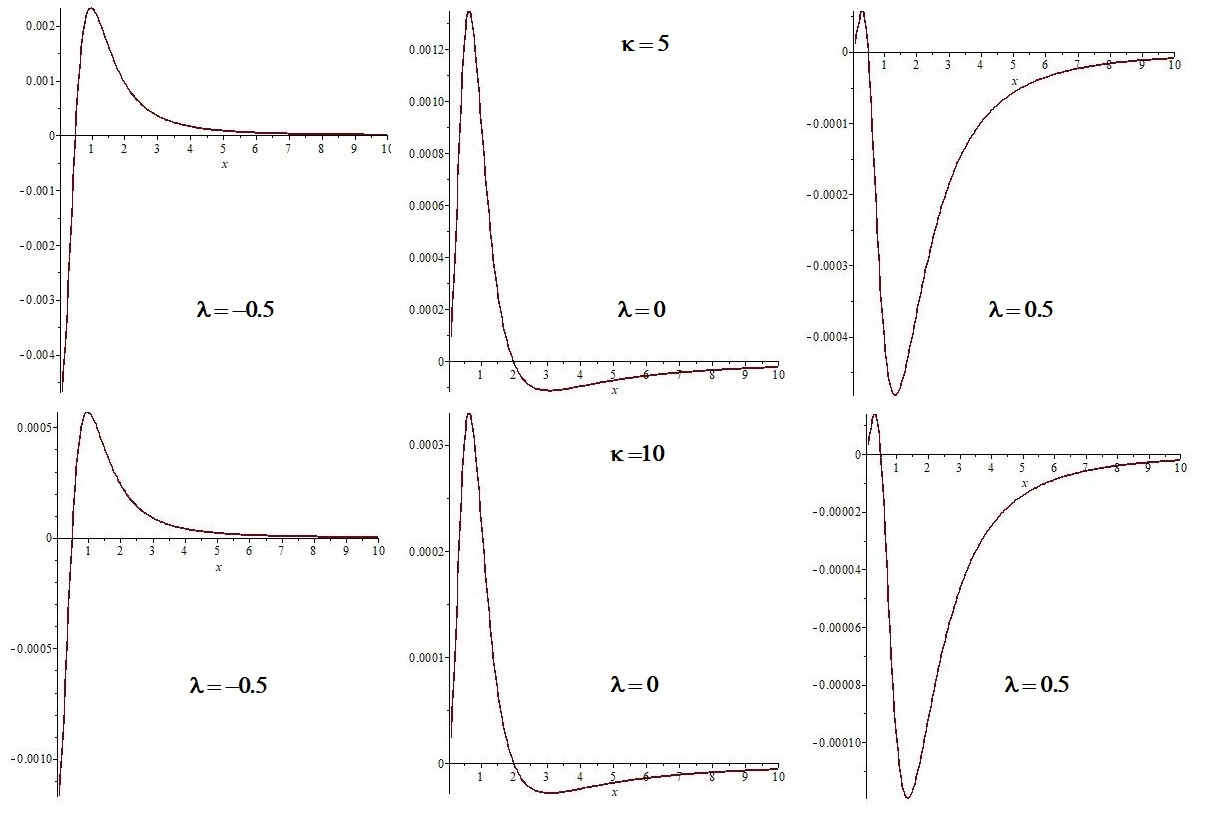}
  \caption{The function ${\cal E}(\kappa,\lambda, x)$ for $\kappa=5$ (upper panels) and $\kappa=10$ (lower panels).}
 \end{figure}}
 
We start with the observation that in our case the variable (\ref{var})  is positively defined without reaching the value $x=0$ if $p\not=0$ such that we can restrict our graphical study to the interval $0.1\leq  x \leq 10$. Then we may ask what it means `large values of $\kappa$' plotting the functions $J$ and ${\cal J}$ on this interval. Thus we see that their graphics tend to approach to each other even for modest values of $\kappa$ as in Fig. (1) where $\kappa=1$ and $\lambda=0,\pm \frac{1}{2}$. For larger values of $\kappa$ (e.g. $\kappa>4$)  the graphics of these functions are overlapping such that we need to resort to the function,  
\begin{equation}
{\cal E}(\kappa,\lambda,x)=1-\frac{{|\cal J|}(\kappa,\lambda,x)|}{|J_{i\kappa+\lambda}(\kappa x)|}\,,
\end{equation}
for pointing out the errors of our approximation.  In Fig. (2) we see how the errors are diminishing when $\kappa$ is increasing from $5$ to $10$.  

The conclusion is that our approximation is numerically satisfactory even for $\lambda \not= 0$.  In practice it is convenient to substitute $\kappa x \to x$ for getting the more homogeneous approximation, 
\begin{equation}\label{unex}
J_{i\kappa+\lambda}(x)\simeq\frac{e^{\frac{\pi\kappa}{2}-\frac{i\pi\lambda}{2}}}{\sqrt{2\pi}}\,\frac{e^{i \sqrt{\kappa^2+x^2}-\frac{i\pi}{4}}}{ (\kappa^2+x^2)^{\frac{1}{4}}}
\left(\frac{\kappa}{x}+\sqrt{1+\frac{\kappa^2}{x^2}}\right)^{-i\kappa-\lambda}\,,
\end{equation}
which is useful in investigating the flat limits of the scalar and spinor fields.

\subsection{Flat limit of the Klein-Gordon field in r.f.v.}

Let us briefly analyze  the flat limits, for $\omega\to 0$, of the mode functions in r.f.v. for $\kappa>\frac{3}{2}$  starting with the time modulation functions (\ref{Ftc})  rewritten, according to Eq. (\ref{IJ}), as 
\begin{equation}\label{Ft}
{\cal F}_p(t)=e^{i\delta_{KG}(p)}\left(\frac{p}{2\omega}\right)^{-i\kappa}\sqrt{\frac{\pi}{\omega}}\frac{e^{\frac{1}{2}\pi\kappa}}{\sqrt{e^{2\pi\kappa}-1}}\,J_{i\kappa}\left(\frac{p}{\omega}e^{-\omega t}\right)\,.
\end{equation}
Here we introduced the auxiliary phase $\delta_{KG}(p)$ we need for removing the pole in $\omega=0$ of the general phase. Note that the second phase factor  assures the correct rest frame limit for $p\to 0$ \cite{CrfvKG}. The flat limit can be evaluated by using the uniform expansion of this Bessel functions (\ref{unex}) where we substitute  $x$  as in Eq. (\ref{var}). Then, according to Eq. (\ref{unex}), we may approximate
\begin{equation}
{\cal F}_p(t_c)\simeq\rho(p,t)e^{i\theta(p,t)}\,,
\end{equation}
where
\begin{eqnarray}
\rho(p,t)&=&\frac{1}{\sqrt{2}(M^2+p^2e^{-2\omega t})^{\frac{1}{4}}}\frac{e^{\pi\kappa}}{\sqrt{e^{2\pi\kappa}-1}}\,,\label{rhoD}\\
\theta(p,t)&=&\delta_{KG}(p)-\frac{\pi}{4}-\frac{M}{\omega} \ln\left(\frac{1}{2\omega}\right)-Mt\nonumber\\
&& -\frac{M}{\omega}\ln\left(M+\sqrt{M^2+p^2e^{-2\omega t}}\right)+\frac{1}{\omega}\sqrt{M^2+p^2e^{-2\omega t}}\,.\label{theta}
\end{eqnarray}
Furthermore, we compute the series of $\theta(p,t)$ around $\omega=0$ where this function has a pole that can be removed by setting 
\begin{equation}\label{deltaKG}
\delta_{KG}(p)=\frac{\pi}{4}+\frac{M}{\omega}\ln \frac{M+\sqrt{M^2+p^2}}{2\omega}-\frac{\sqrt{M^2+p^2}}{\omega}\,.
\end{equation} 
In this manner we generate the phase factor,
\begin{equation}
e^{i\delta_{KG}(p)}=e^{\frac{i \pi}{4}}\left(\frac{M+\sqrt{M^2+p^2}}{2\omega} \right)^{\frac{iM}{\omega}}e^{-i\frac{\sqrt{M^2+p^2}}{\omega}}\,,
\end{equation}
without physical significance, but necessary for deriving the convenient approximation that can be used in applications,
\begin{equation}\label{Ffin}
{\cal F}_p(t)\simeq\frac{e^{\pi\kappa}}{\sqrt{e^{2\pi\kappa}-1}}\frac{e^{i\theta_{KG}(p,t)}}{\sqrt{2}(M^2+p^2e^{-2\omega t})^{\frac{1}{4}}}\,.  
\end{equation}
The regularized phase, 
\begin{eqnarray}
\theta_{KG}(p,t)&=&-Mt
 -\frac{M}{\omega}\ln\left(\frac{M+\sqrt{M^2+p^2e^{-2\omega t}}}{M+\sqrt{M^2+p^2}}\right)\nonumber\\
&+&\frac{1}{\omega}\left(\sqrt{M^2+p^2e^{-2\omega t}}-\sqrt{M^2+p^2}\right),.\label{thetaKG}
\end{eqnarray}
is obtained by substituting  in Eq. (\ref{theta}) the phase $\delta_{KG}$ defined by Eq. (\ref{deltaKG}). For small values of $\omega$  we may use the Taylor series 
\begin{equation}\label{thetaKGex}
\theta_{KG}(p,t)=-\sqrt{M^2+p^2}\,t+\frac{\omega p^2 t^2}{2\sqrt{M^2+p^2}}-\frac{\omega^2 p^2(2M^2+p^2)t^3}{6(M^2+p^2)^\frac{3}{2}}+{\cal O}(\omega^3)\,.
\end{equation}
finding that in the flat limit, when $\lim_{\omega\to 0}M=m$, and
\begin{equation}
\lim_{\omega\to 0}{\cal F}_p(t)=\frac{e^{-iE(p)t}}{\sqrt{2E(p)}}\,, \quad E(p)=\sqrt{m^2+p^2}\,,
\end{equation}
we recover just the Minkowkian time modulation functions. 

\subsection{Flat limit of the Dirac field in r.f.v.}

For analyzing how the flat limit can be reached in the case of the Dirac field it is convenient to rewrite the time modulation functions (\ref{uuI})  in the chart $\{t,\vec{x}\}$ as
\begin{equation}
u_p^{\pm}(t)=\pm e^{i\delta_D(p)\pm \frac{i\pi}{4}} p^{-i\kappa}\frac{e^{\frac{\pi\kappa}{2}}}{\sqrt{e^{2\pi \kappa}+1}}\sqrt{\frac{\pi}{\omega}\, p e^{-\omega t}}\, J_{i\kappa\mp\frac{1}{2}}\left(\frac{p}{\omega}e^{-\omega t}\right)\,,
\end{equation}
after introducing the phase $\delta_D(p)$ which should take over the singularities of the general phase as in the previous case. The uniform expansion (\ref{unex}) with $\lambda=\pm\frac{1}{2}$  helps us to approximate
\begin{equation}
u_p^{\pm}(t)\simeq\rho^{\pm}(p,t) e^{i\theta(p,t)}\,,
\end{equation} 
where
\begin{eqnarray}
\rho^+(p,t)&=&\frac{\sqrt{\sqrt{m^2+p^2e^{-2\omega t}}+m}}{\sqrt{2}(m^2+p^2 e^{-2\omega t})^{\frac{1}{4}}}\frac{e^{\pi\kappa}}{\sqrt{e^{2\pi \kappa}+1}}\,,\label{RP}\\
\rho^-(p,t)&=&\frac{p e^{-\omega t}}{\sqrt{2}(m^2+p^2 e^{-2\omega t})^{\frac{1}{4}}\sqrt{\sqrt{m^2+p^2e^{-2\omega t}}+m}}\frac{e^{\pi\kappa}}{\sqrt{e^{2\pi \kappa}+1}}\,,\label{RM}\\
\theta(p,t)&=&\delta_D(p)+\frac{\pi}{4}-mt\nonumber\\
&& -\frac{m}{\omega}\ln\left(m+\sqrt{m^2+p^2e^{-2\omega t}}\right)+\frac{1}{\omega}\sqrt{m^2+p^2e^{-2\omega t}}\,.\label{thetaD}
\end{eqnarray}
We observe that  the obvious identity 
\begin{equation}
pe^{-\omega t}=\sqrt{\sqrt{m^2+p^2e^{-2\omega t}}+m}\,\sqrt{\sqrt{m^2+p^2e^{-2\omega t}}-m}
\end{equation}
can be substituted in Eq. (\ref{RM}) for getting a more symmetric and compact form.
Furthermore, we expand the function $\theta(p,t)$ around $\omega=0$ and we chose
\begin{equation}
\delta_D(p)=-\frac{\pi}{4}+\frac{m}{\omega}\ln(\sqrt{m^2+p^2}+m)-\frac{1}{\omega}\sqrt{m^2+p^2}\,,
\end{equation}
for eliminating the effects of the pole in $\omega=0$.  Thus  we arrive at the final expansions for large values of $\kappa$ (when $\omega \to 0$) that reads
\begin{equation}\label{uaprox}
u_p^{\pm}(t)\simeq \frac{e^{\frac{\pi m}{\omega}}}{\sqrt{e^{2 \frac{\pi m}{\omega}}+1}}\frac{\sqrt{\sqrt{m^2+p^2e^{-2\omega t}}\pm m}}{\sqrt{2}(m^2+p^2 e^{-2\omega t})^{\frac{1}{4}}}\, e^{i\theta_D(p,t)}\,.
\end{equation}
The regularized phase,  
\begin{eqnarray}
\theta_D(p,t)&=&-m t  -\frac{m}{\omega}\ln\left(\frac{m+\sqrt{m^2+p^2e^{-2\omega t}}}{m+\sqrt{m^2+p^2}}\right)\nonumber\\
&+&\frac{1}{\omega}\left(\sqrt{m^2+p^2e^{-2\omega t}} -\sqrt{m^2+p^2}\right)\,,
\end{eqnarray}
obtained after substituting $\delta_D$ in Eq. (\ref{thetaD}), can be expanded as 
\begin{equation}\label{thetaD1}
\theta_{D}(p,t)=-\sqrt{m^2+p^2}\,t+\frac{\omega p^2 t^2}{2\sqrt{m^2+p^2}}-\frac{\omega^2 p^2(2m^2+p^2)t^3}{6(m^2+p^2)^\frac{3}{2}}+{\cal O}(\omega^3)\,,
\end{equation}
laying out a similar form as the phase (\ref{thetaKG}) of the scalar field but with the usual mass $m$ instead of the dynamical one.

Finally we verify that in the flat limit we obtain the usual Minkowskian time modulation functions
\begin{equation}
\lim_{\omega\to 0}u_p^{\pm}(t) =\sqrt{\frac{E(p)\pm m}{2 E(p)}}\, e^{-iE(p)t}\,. 
\end{equation}

\subsection{Physical consequences}

Solving the problem of the flat limit we derived the suitable phases that complete the phases we introduced previously for defining the r.f.v.. We obtain thus the final form the scalar time modulation functions can be written now as 
\begin{equation}\label{Fdef}
{\cal F}_p(t)=e^{i\alpha_{KG}(p)}\sqrt{\frac{\pi}{\omega}}\frac{e^{\frac{1}{2}\pi\kappa}}{\sqrt{e^{2\pi\kappa}-1}}\,J_{i\kappa}\left(\frac{p}{\omega}e^{-\omega t}\right)\,,
\end{equation}
where $\kappa=\frac{M}{\omega}$ while the global phase,
\begin{eqnarray}
\alpha_{KG}(p)&=&\delta_{KG}(p)-\kappa \ln\left(\frac{p}{2\omega}\right) \nonumber\\
&=&\frac{\pi}{4}+\frac{M}{\omega}\ln \frac{M+\sqrt{M^2+p^2}}{p}-\frac{\sqrt{M^2+p^2}}{\omega}\,,
\end{eqnarray}
depends on the dynamical mass ({\ref{Mko}).

For the Dirac time modulation functions we may write a similar result,
 \begin{equation}\label{ufin}
u_p^{\pm}(t)=\pm e^{i\alpha_D(p)\pm \frac{i\pi}{4}} \frac{e^{\frac{\pi\kappa}{2}}}{\sqrt{e^{2\pi \kappa}+1}}\sqrt{\frac{\pi}{\omega}\, p e^{-\omega t}}\, J_{i\kappa\mp\frac{1}{2}}\left(\frac{p}{\omega}e^{-\omega t}\right)\,,
\end{equation}
where now $\kappa=\frac{m}{\omega}$ and
\begin{eqnarray}
\alpha_D(p)&=&\delta_{D}(p)-\kappa \ln p\nonumber\\
&=&-\frac{\pi}{4}+\frac{m}{\omega}\ln\left(\frac{\sqrt{m^2+p^2}+m}{p}\right)-\frac{\sqrt{m^2+p^2}}{\omega}\,,
\end{eqnarray}
is very similar with the scalar phase but with the genuine mass $m$ instead of the dynamical one,  $M$.

Why the phases are so important in the de Sitter spacetime as long as these do not affect the scalar products and do not contribute to the expressions of the transition probabilities. A specific feature of the de Sitter QFT is that the forms of some one-particle operators including the Hamiltonian (or energy) one are strongly dependent on  the phases which are functions of $p$.  We remind the reader that, after canonical quantization, the one-particle Hamiltonian operator of any quantum field,  $\Psi$,   can be calculated by using the corresponding relativistic scalar product,  ${\cal H}=:\langle \Psi, H\Psi\rangle:$, in which we respect the normal ordering of the field operators \cite{BDR}. The energy operator,
\begin{equation}
H=i\partial_t+\omega \vec{x}\cdot\vec{P}=i\partial_t-i\omega \vec{x}\cdot \nabla
\end{equation}
is the same for any free field since it does not have spin parts \cite{nach,CGRG}. 

In the case of the Klein-Gordon field we consider the normalized mode functions (\ref{fp}) with the time modulation functions  (\ref{Fdef}). Then it is not difficult to verify the identity 
\begin{equation}\label{H}
(Hf_{\vec{p}})=\left[-i\omega \left(p^i\partial_{p_i}+{\frac{3}{2}}\right)-\omega p^i\partial_{p^i}\alpha_{KG}(p)\right]f_{\vec{p}}
\end{equation}
which allows us to derive the form of the one-particle Hamiltonian operator  as,
\begin{eqnarray}
{\cal H}_{KG}&=&:\langle \Phi, H\Phi\rangle_{KG}:\,= \int d^3 p\, \sqrt{M^2+p^2}
\left[a^{\dagger}({\bf p}) a({\bf p})
+{b}^{\dagger}({\bf p}){b}({\bf p})\right]\nonumber\\
&&~~~+\frac{i\omega}{2}\int d^3p\, p^i \left\{ \left[\, a^{\dagger}({\bf
p})\stackrel{\leftrightarrow}{\partial}_{p_i} a({\bf p})\right]+ \left[\,
b^{\dagger}({\bf p}) \stackrel{\leftrightarrow}{\partial}_{p_i} b({\bf
p})\right]\right\}\,,\label{Ham}
\end{eqnarray}
since
\begin{equation}\label{Eap}
\omega p^i\partial_{p^i}\alpha_{KG}(p)=-\sqrt{M^2+p^2}\,.
\end{equation}
We have thus the nice surprise to see that by fixing the correct phases requested by the rest and flat limits we obtain an operator whose flat limit,
\begin{equation}
\lim_{\omega \to 0} {\cal H}_{KG}= \int d^3 p\, \sqrt{m^2+p^2}
\left[a^{\dagger}({\bf p}) a({\bf p})
+{b}^{\dagger}({\bf p}){b}({\bf p})\right]\,,
\end{equation}
is just the well-known Hamiltonian operator of the Minkowskian QFT. A similar result can be obtained for the Dirac field. Therefore, the flat limit of the entire de Sitter QFT is the just the QFT of special relativity. 

\section{Concluding remarks}

We derived the definitive forms of the fundamental solutions of the Klein-Gordon and Dirac fields whose frequencies are separated in the rest frames as in special relativity having, in addition,  Minkowskian  flat limits.  

Similar results concerning the phase factors or mode expansions as the second term of Eq. (\ref{Ham}) were obtained in premiere for the scalar and spinor fields long time ago in Ref. \cite{nach} where the adiabatic vacuum was considered. Subsequent studies refined these results  \cite{born,CPasc,CD} such that now we can conclude  that the regularizes phases derived so far are very similar to those obtained here in r.f.v.. This is because in the adiabatic vacuum, where the rest limits are undefined, one forced some {\em ad hoc} changes of  the time modulation functions, introducing phase factors proportional to $p^{-i\kappa}$ similar to those arising naturally in r.f.v..  

However, apart from the known regularized phases recovered here,  we report  new results as the final form of the time modulation functions (\ref{Fdef}) and (\ref{ufin}) in r.f.v. and the approximations (\ref{Ffin}) and (\ref{uaprox}) that can be used in applications for deriving transition amplitudes between states defined in this vacuum.
 
The final conclusion is that  the rest and flat limits appear as the cornerstones of the QFT on the de Sitter expanding universe determining the form of the time modulation functions in a natural manner and  setting thus the structure of the one-particle operators. In our opinion, the results presented here are an argument that the r.f.v. could be universal.

\subsection*{Acknowledgments}

This work is partially supported by a grant of  the Romanian Ministry of Research and Innovation, CCCDI-UEFISCDI, project number  PN-III-P1-1.2-PCCDI-2017-0371. 

\appendix

\setcounter{section}{0}\renewcommand{\thesection}{\Alph{section}}
\setcounter{equation}{0} \renewcommand{\theequation}
{A.\arabic{equation}}
\section{Some properties of Bessel functions}

The modified Bessel functions $I_{\nu}(z)$ and $K_{\nu}(z)$ are related as \cite{NIST}
\begin{eqnarray}
K_{\nu}(z)&=&K_{-\nu}(z)=\frac{\pi}{2}\frac{I_{-\nu}(z)-I_{\nu}(z)}{\sin\pi \nu}\,,\label{IK}\\
I_{\pm\nu}(z)&=&e^{\mp i\pi\nu}I_{\pm\nu}(-z)\nonumber\\
&=&\frac{i}{\pi}\left[K_{\nu}(-z)-e^{\mp i\pi\nu}K_{\nu}(z)\right]\,.\label{KI}
\end{eqnarray}
Their Wronskians  give  the identities we need for normalizing the mode functions. For $\nu=i\kappa$ we obtain
\begin{equation}
i I_{i\kappa}(i s) \stackrel{\leftrightarrow}{\partial_{s}}I_{-i\kappa}(is)= \frac{2\, {\rm sinh}\,\pi\kappa}{\pi s}\,,\label{IuIu} 
\end{equation}
while the identity
\begin{equation}
i K_{\nu}(-i s) \stackrel{\leftrightarrow}{\partial_{s}}K_{\nu}(is)
=\frac{\pi}{|s|}\,,\label{KuKu}
\end{equation}
holds for any $\nu$.

For $|z|\to \infty$ and any $\nu$ we have,
\begin{equation}\label{Km0}
 I_{\nu}(z) \to \sqrt{\frac{\pi}{2z}}e^{z}\,, \quad K_{\nu}(z) \to K_{\frac{1}{2}}(z)=\sqrt{\frac{\pi}{2z}}e^{-z}\,.
\end{equation} 
In the limit of $|z|\to 0$ the functions $I_{\nu}$ behave as  
\begin{equation}\label{I0}
I_{\nu}(z)\sim \frac{1}{\Gamma(\nu+1)} \left(\frac{z}{2}\right)^{\nu}\,,
\end{equation}
while for the functions $K_{\nu}$ we have to use Eq. (\ref{IK}).

The modified Bessel functions $I$ are related to the usual ones as 
\begin{equation}\label{IJ}
I_{\nu}(-ix)=e^{-\frac{i\pi\nu}{2}}J_{\nu}\left(x\right)\,,\quad \forall\, x\in{\Bbb R},\, \nu\in{\Bbb C}\,,
\end{equation}
for which we can apply the following uniform asymptotic expansion \cite{bat}, 
\begin{eqnarray}
J_{i\kappa}(\kappa x) &=&\frac{e^{\frac{\pi\kappa}{2}}}{\sqrt{2\kappa}\,\pi}\,\frac{e^{i \kappa\sqrt{1+x^2}-\frac{i\pi}{4}}}{ (1+x^2)^{\frac{1}{4}}}
\left(\frac{1}{x}+\sqrt{1+\frac{1}{x^2}}\right)^{-i\kappa}\nonumber\\
&\times&\left[\sum_{n=0}^{n=N} \frac{(2 i)^n\Gamma(n+\frac{1}{2})}{\kappa^n (1+x^2)^{\frac{n}{2}}}\,a_n (x)+ {\cal O}(\kappa^{-N-1})\right]\,,\label{unexA}
\end{eqnarray}
where $a_n$ are polynomials in $(1+x^2)^{-1}$, 
\begin{eqnarray}
&&a_0(x)=1\,, \quad a_1(x)=-\frac{1}{8}+\frac{5}{24}(1+x^2)^{-1}\,,\nonumber\\
&& a_2(x)=\frac{3}{128}-\frac{77}{576}(1+x^2)^{-1}+\frac{385}{3456}(1+x^2)^{-2},...\nonumber
\end{eqnarray}
with coefficients less that 1 that do not depend  on $\kappa$. This expansion holds for any $\kappa, x >0$.  Here we consider the case of $N=0$ in which the contribution of the above sum reduces to $\Gamma(\frac{1}{2})=\sqrt{\pi}$.

\end{document}